\documentclass[twocolumn]{article}
\usepackage[utf8]{inputenc}
\usepackage[dvipdfmx]{graphicx}
\usepackage{siunitx}
\usepackage{authblk}
\usepackage{hyperref}
\usepackage{amsmath}
\usepackage{bm}

\hypersetup{
    colorlinks=true,
    linkcolor=blue,
    filecolor=magenta,      
    urlcolor=cyan,
    }

\usepackage[margin=2.5cm]{geometry}

\title{\vspace{-15mm} Background oriented schlieren technique  with\\ fast Fourier demodulation \\ for measuring large density-gradient fields of fluids}

\author[1]{\small{Takaaki Shimazaki}}
\author[1]{Sayaka Ichihara}
\author[1,2]{Yoshiyuki Tagawa}

\affil[1]{Department of Mechanical Systems Engineering, Tokyo University of Agriculture and Technology, Koganei Campus 6-204, 2-24-16 Nakacho, Koganei, Tokyo, Japan
}
\affil[2]{Institute of Global Innovation Research, Tokyo University of Agriculture and Technology, Koganei Campus 6-204, 2-24-16 Nakacho, Koganei, Tokyo, Japan
}

\date{}

\begin{document}

\twocolumn[

\maketitle

\vspace{-10mm}

\paragraph{Abstract} In order to measure the large density-gradient fields of fluids such as underwater shock waves, we employed a fast Fourier demodulation called Fast Checkerboard Demodulation (FCD) method in Background Oriented Schlieren (BOS) technique.
BOS is a simple image-based measurement technique that detects the apparent displacement (local distortion) of a background image caused by the density gradient of the fluid in front of the background.
The cross-correlation particle image velocimetry (PIV) method, which is commonly employed in the BOS technique for detecting the apparent displacement, uses a random-dot background, whereas FCD uses a periodic pattern as the background (e.g., checkerboard pattern, lattice grid pattern (grid scale)) to detect the apparent displacement as the phase change of the pattern in the Fourier space.
In this study, we measured the apparent displacement, which is proportional to density gradient of fluid, of an underwater shock wave using FCD and PIV. 
The results showed that FCD can measure a displacement gradient of up to 2.5 times larger than that which PIV can measure.
Furthermore, we systematically investigated the measurement limit of FCD-BOS by changing several parameters of the periodic patterns, such as the grid size.
Also, we explored the related parameters in the Fourier space to understand the limitations.
It is worth noting that FCD-BOS with lattice grid pattern (grid scale) can measure the apparent displacement as accurately as that with a custom-made pattern, indicating that large density-gradient fields of fluids can be measured using a simple setup with a commercially available (inexpensive) pattern.

\vspace{20pt}

]

\section{Introduction}

Background Oriented Schlieren (BOS) technique is a non-contact method for measuring the density field in a fluid \cite{Dalziel2000,Raffel2000,Richard2001,Meier2002,Venkatakrishnan2004}.
Compared to conventional Schlieren imagin and optical interferometry, the measurement system is quite simple:
it consists of only a background, camera, and light source.
BOS technique has been applied to various density fields, such as gas flows \cite{Ting2013,Vinnichenko2014,Rajshekhar2018,Vinnichenko2019,Hendiger2020}, shock waves \cite{Meier2002,Venkatakrishnan2004,Yamamoto2015,Hayasaka2016}, flames \cite{veser2011,Su2016,Grauer2018,Michalski2018,Hayasaka2019}, and liquid-gas interfaces \cite{Gupta2019,Vinnichenko2020}.
Furthermore, it has also been applied to various length scales, from micrometers \cite{Raffel2000,Yamamoto2015,Hayasaka2016,kaneko2021background} to meters \cite{Venkatakrishnan2004,Hargather2013}.
Moreover, complex 3D density fields have been also measured using multi-directional BOS imaging \cite{Nicolas2017,Grauer2018,Liu2020}.

\begin{figure}[t]
\begin{center}
\includegraphics[width=1.0\columnwidth]{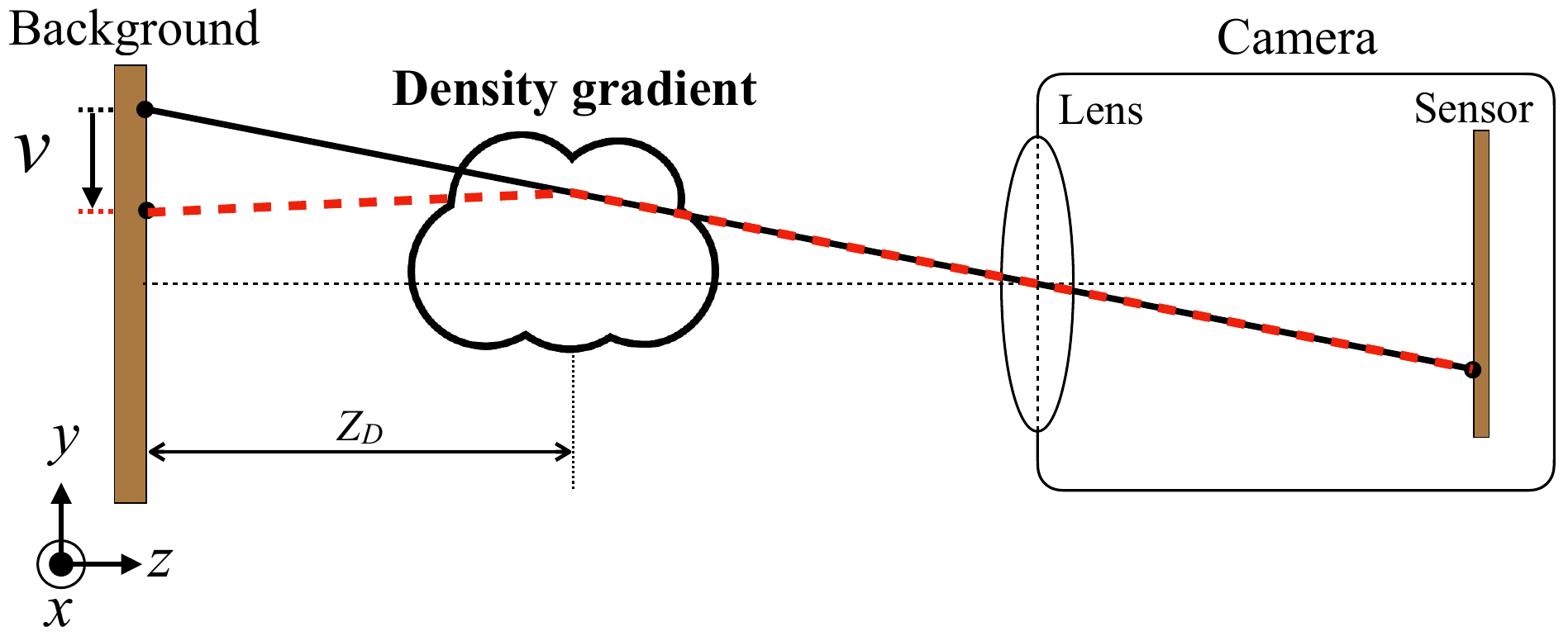}
\caption{Illustration of BOS technique. The black and red lines represent light paths (optical path lengths) with and without density gradient, respectively.}
\label{fig:bos}
\end{center}
\end{figure}

\begin{figure}[t]
\begin{center}
\includegraphics[width=0.9\columnwidth]{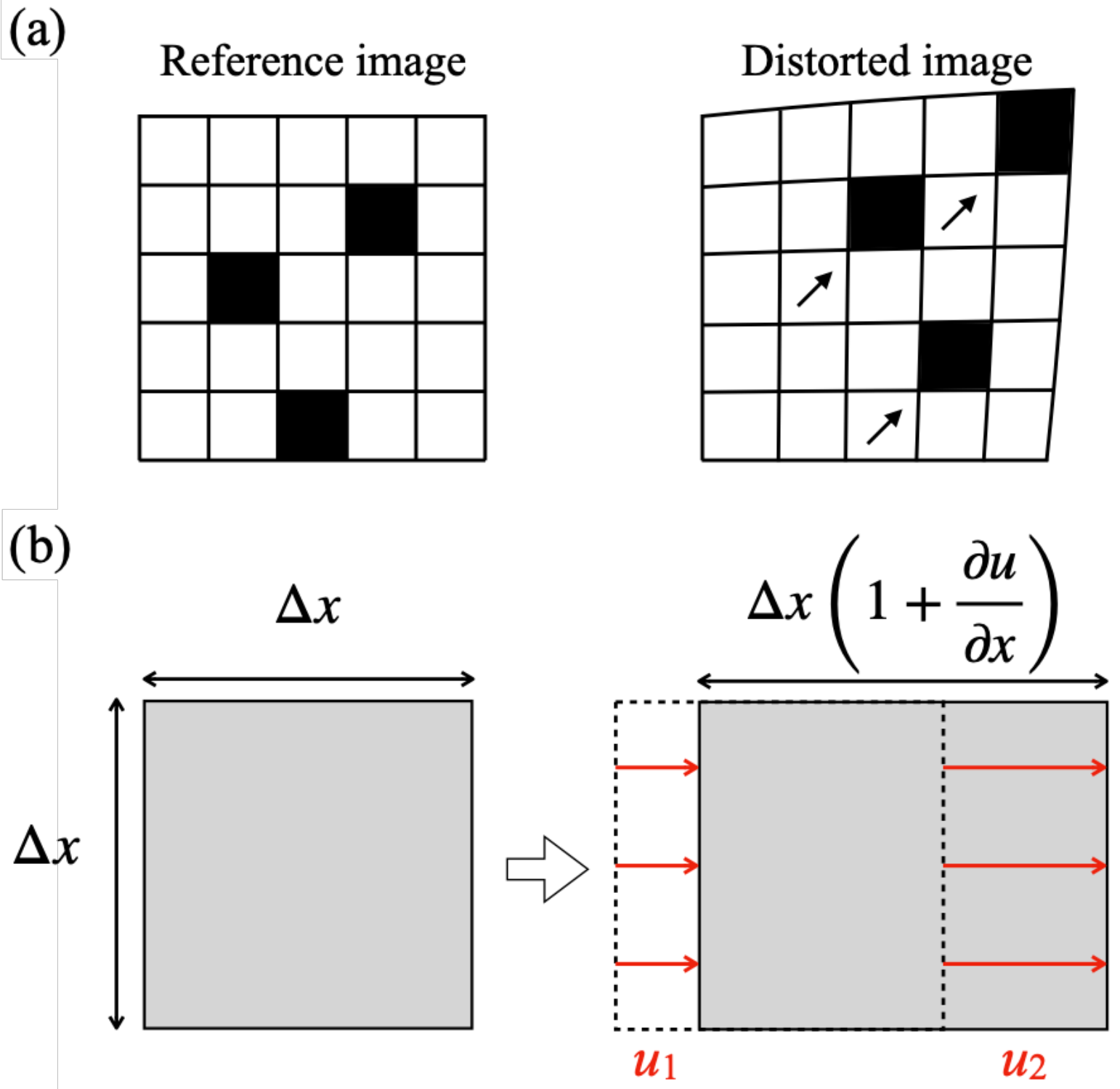}
\caption{(a) (Left) A background image of dot pattern without apparent displacement (Reference image). (Right) A background image with apparent displacement (Distorted image). Black squares are not only translated but also deformed, causing problems to the displacement detection by PIV. (b) Simple situation under a lateral apparent displacements $u_1$ and $u_2$. The magnitude of the dot deformation is represented by the displacement gradient $\partial u/ \partial x$.}
\label{fig:deformation}
\end{center}
\end{figure}

Figure \ref{fig:bos} shows the illustration of BOS technique.
BOS technique quantifies the density-gradient field in front of a background by comparing two background images with and without the density gradient which are captured using a camera.
The background image with density gradient looks distorted as compared to the background image without the density gradient because the refractive index of a fluid changes with density.
Such distortion is quantified as ''apparent displacement".
It is extremely important for the BOS technique to accurately measure the apparent displacement between the background images \cite{Goldhahn2007, Atcheson2009, Hargather2012}.
A random-dot background is commonly used to obtain apparent displacement of images through analysis by digital image correlation (DIC), which is known as particle image velocimetry (PIV) in the fluid dynamics research community \cite{Keane1992,Westerweel1997,Prasad2000,Meinhart2000} or optical flow \cite{Liu2008,Liu2015,Wang2015}.
However, neither PIV nor optical flow is capable of detecting displacements of large density gradients because 
they assume parallel displacement of the background image, while in reality, random dots experience a shape distortion due to large density gradients as illustrated in Figure \ref{fig:deformation}(a).

Let us consider a simple situation in which a square ${\rm \Delta} x \times {\rm \Delta} x$ on a background is distorted in one direction (Figure \ref{fig:deformation}(b)).
Because the magnitudes of the displacements $u_1$ on the left side of the square and $u_2$  on the right side of the square are different, the size of the square after distortion is expressed as ${\rm \Delta} x'={\rm \Delta} x+u_2-u_1$, and the rate of shape distortion is $({\rm \Delta} x'-{\rm \Delta} x)/{\rm \Delta} x = (u_1-u_2)/{\rm \Delta} x \approx \partial u/ \partial x$.
Thus, the magnitude of the distortion depends on the displacement gradient $\partial u/ \partial x$, i.e., the second-order spatial differential of the density field of the fluid.
When the displacement gradient is too large, PIV cannot properly acquire the apparent displacement \cite{Moisy2009}.
Hence, there is a limitation in the existing PIV-based BOS measurements for underwater shock waves and flames \cite{Elbaz2014} due to their large displacement gradients.

In this study, to solve this problem fundamentally, we employed the Fast Checkerboard Demodulation (FCD) method which was proposed by Wildeman \cite{Wildeman2018}, because unlike PIV, FCD can measure the displacement even when the background pattern is distorted.
FCD uses a background with a two-dimensional periodic pattern to detect the displacement field from the change of wavenumber in the Fourier space. 
By using FCD, Wildeman \cite{Wildeman2018} has successfully detected waves on a water surface. 
This study takes advantage on the robustness of FCD in detecting large apparent displacements.
It is the first time that FCD method is used in a BOS technique.
In our measurement system, the background patterns of horizontal and vertical lines (lattice grid pattern) and of a black-and-white square pattern (checker pattern) are used. 
For comparison, a density-gradient field was measured by both PIV and FCD.
Also, the data measured using hydrophone was used as the reference (correct) data. 
We measured underwater shock waves since they are good examples of flow field with large displacement gradients.
We also performed numerical validation by adding the displacement to a synthetic background image to gain a better understanding on the experimental values and the measurement limits of FCD-BOS.

This paper is structured as follows.
Principles of BOS technique and FCD method are described in Section 2. 
The experimental setup and results are explained in Section 3.
We compared the apparent displacement and displacement gradient obtained using FCD and PIV with those estimated from the pressure measured by the hydrophone.
Finally, we also analyzed the synthetic image under ideal conditions for discussion on the measurement limit of FCD-BOS in Section 4.


\section{Principle}

\subsection{Background Oriented Schlieren (BOS) technique}

BOS is a technique which visualizes and quantifies the density gradient by detecting variations in the refractive index of the fluid.
As shown by the illustration in Figure \ref{fig:bos}, the measurement target is placed between a camera and a background where the camera is focused onto the background to capture images with and without density gradients.
FCD or PIV is then used to determine the apparent displacement by comparing both images.
The apparent displacement in $y$-direction, $v$, (Figure \ref{fig:bos}) can be related to the gradient of the refractive index as shown in the following equation \cite{Venkatakrishnan2004} \textcolor{blue}{:}
\begin{equation}
v = \frac{Z_D}{n_0}\int \frac{\partial n}{\partial y}dz,
\label{eq:v_n}
\end{equation}
where $Z_D$ is the distance from the background to the center of the density gradient, $n$ is the refractive index of the measurement target, and $n_0$ is the refractive index of the ambient fluid.
The relationship between the refractive index $n$ and the fluid density $\rho$ is given by the Gladstone-Dale equation: 
\begin{equation}
n=K\rho+1,
\label{eq:gladstone}
\end{equation}
where $K$ is the Gladstone-Dale constant, which is $3.34\times10^{-4}$ $\rm{m^3/kg}$ for water \cite{Venkatakrishnan2004}.
Substituting Eq. \ref{eq:gladstone} into Eq. \ref{eq:v_n} leads to
\begin{equation}
v=\frac{K Z_D}{K \rho_0 + 1} \int \frac{\partial \rho}{\partial y} dz,
\label{eq:v_rho}
\end{equation}
where $\rho_0$ is the density of the ambient fluid.
Similarly, the apparent displacement in $x$-direction, $u$, is
\begin{equation}
u=\frac{K Z_D}{K \rho_0 + 1} \int \frac{\partial \rho}{\partial x} dz.
\label{eq:u_rho}
\end{equation}

In this study, the apparent displacement measured by the BOS technique and that estimated from direct measurement of pressure using a hydrophone are compared to validate the ability of FCD-BOS in detecting apparent displacement.
The procedure for estimating the apparent displacement from the pressure measured by the hydrophone is explained in $\S$ 2.3

\subsection{Fast Fourier demodulation of a periodic background}

%
\begin{figure*}[!ht]
\begin{center}
\includegraphics[width=2\columnwidth]{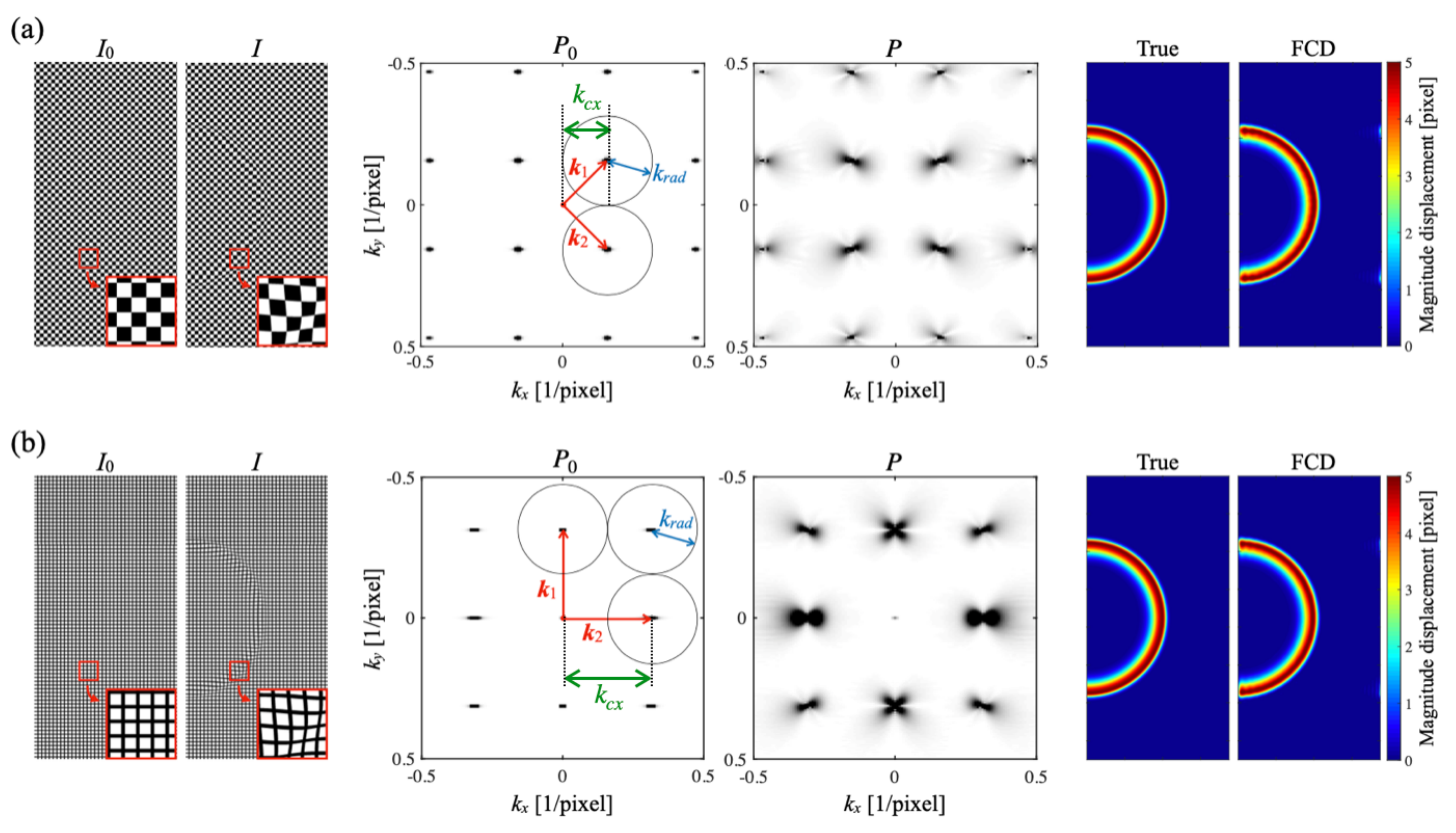}
\caption{(a) The checkered pattern with tiles of 20 \si{\um} $\times$ 20 \si{\um}. (b) The lattice grid pattern with black lines of 5 \si{\um} wide and white regions of 15 \si{\um} $\times$ 15 \si{\um}. A distorted image $I$ (1500 $\times$ 750 pixels) was created by displacing the reference image $I_0$. $P_0$ and $P$ are the power spectra in the Fourier space ($k$-space) of $I_0$ and $I$, respectively. $\bm k_1$ and $\bm k_2$ are the carrier peaks. The radius of a circle in $P_0$ is $k_{rad}$. $k_{cx}$ is the norm of carrier peak $k_c$ along $x$-axis. The synthetic displacement (true) and the displacement detected by the FCD are shown.}
\label{fig:FCD}
\end{center}
\end{figure*}
A spatial Fourier demodulation technique of a periodic background, which is known as Fast Checkerboard Demodulation (FCD) method and was proposed by Wildeman \cite{Wildeman2018}, allows the quantitative extraction of the apparent displacement field when a periodic checkered background is used.
By photographing a background with a periodic pattern (a lattice grid pattern or a checkered pattern in this study) and performing Fourier transform, the apparent displacement on the background is obtained as the phase change.
Reference image $I_0(\bm{\xi})$ is expressed as a general 2D periodic pattern as
\begin{equation}
I_0({\bm \xi}) = \sum_{m=-\infty}^\infty \sum_{n=-\infty}^\infty a_{mn}e^{i(m{\bm k_1}+n{\bm k_2}) \cdot {\bm \xi}},
\end{equation}
where $\bm \xi$ denotes the pixel coordinates ($x$, $y$) in the image, $\bm k_1$ and $\bm k_2$ denote the orthogonal wavenumber vectors extracted from the reference image (called carrier peaks), and $a_{mn}$ denotes the Fourier coefficient.
The distorted image $I(\bm \xi)$ is the result when the reference image $I_0$ is modulated by the apparent displacement ${\bm u}({\bm \xi})$, as follows:
\begin{equation}
I(\bm{\xi})=I_0({\bm \xi} - {\bm u}({\bm \xi})).
\end{equation}
By performing Fourier transforming on $I_0$ and $I$, we obtain the respective power spectras $P_0$ and $P$ with peaks at $m \bm k_1+n \bm k_2$, which is expressed using integers $m,n$, as shown in Figure \ref{fig:FCD}.
By extracting the signals within a radius of $k_{rad}$ centered from $\bm k_1$ and $\bm k_2$ in $P_0$ and $P$, respectively, and performing the inverse Fourier transform, we obtain
\begin{gather}
g_1({\bm \xi})=a_{c1} e^{i{\bm k_1} \cdot {\bm \xi}}, \\
g_2({\bm \xi})=a_{c2} e^{i{\bm k_2} \cdot {\bm \xi}}, \\
g_1'({\bm \xi})=a_{c1}' e^{i{\bm k_1} \cdot ({\bm \xi - \bm u(\bm \xi)})}, \\
g_2'({\bm \xi})=a_{c2}' e^{i{\bm k_2} \cdot ({\bm \xi - \bm u(\bm \xi)})},
\end{gather}
where $a_{c1}, a_{c2}, a_{c1}'$, and $\ a_{c2}'$ are coefficients derived from $a_{mn}$.
The phase fields $\phi_1(\bm \xi), \phi_2(\bm \xi)$ extracted from two linearly independent carrier peaks are
\begin{gather}
\label{eq:phi1} \phi_1(\bm \xi) = {\rm Im} ( \ln(g_1 g_1'^* ) ) = {\bm k_1} \cdot {\bm u}({\bm \xi}) \\
\label{eq:phi2} \phi_2(\bm \xi) = {\rm Im} ( \ln(g_2 g_2'^* ) ) = {\bm k_2} \cdot {\bm u}({\bm \xi})
\end{gather}

or 

\begin{equation}
\left(
    \begin{array}{ccc}
      \phi_1 \\
      \phi_2
    \end{array}
  \right)
= \left(
    \begin{array}{ccc}
      k_{1x} & k_{1y} \\
      k_{2x} & k_{2y}
    \end{array}
  \right)
\left(
    \begin{array}{ccc}
      u \\
      v
    \end{array}
  \right),
\label{eq:phi_2_2}
\end{equation}
where $k_{1x}$ is the $x$-component of vector $\bm k_1$, $k_{1y}$ is the $y$-component of vector $\bm k_1$, $k_{2x}$ is the $x$-component of vector $\bm k_2$, $k_{2y}$ is the $y$-component of vector $\bm k_2$, and $\bm u(\bm \xi)=(u,v)$. 
By solving Eq. \ref{eq:phi_2_2}, we obtain the displacement $u, v$ directly as
\begin{equation}
\left(
    \begin{array}{ccc}
      u \\
      v
    \end{array}
  \right)
=  \dfrac{1}{k_{1x} k_{2y} - k_{1y} k_{2x}}
\left(
    \begin{array}{ccc}
      k_{2y} & -k_{1y} \\
      -k_{2x} & k_{1x}
    \end{array}
  \right)
\left(
    \begin{array}{ccc}
      \phi_1 \\
      \phi_2
    \end{array}
  \right).
\end{equation}

To avoid aliasing and phase wrapping, following three criteria must be satisfied for the displacement measurement \cite{Wildeman2018}:
\begin{gather}
\label{eq:fcd1} k_s < k_{rad}, \\
\label{eq:fcd2} k_c |\bm u'| < k_{rad}, \\
\label{eq:fcd3} k_c |\bm u| < \pi,
\end{gather}
where $k_s$ is the wavenumber of the displacement, $k_c$ ( ${k_c} = |\bm{k_{1}}| = |\bm{k_{2}}|$ ) is the norm of the carrier peak of the reference image, and $\bm u'$ is the displacement gradient.
In this study, we focus on the value on $x$-axis.
For this reason,  although Wildeman \cite{Wildeman2018} uses $k_c$  in the criteria, we utilized the norm of carrier peak along $x$-axis, i.e., $k_{cx}$. 
Thus, the criteria correspond to Eq. \ref{eq:fcd2} and Eq. \ref{eq:fcd3} in this study are
\begin{gather}
\label{eq:fcd2_2} k_{cx} |\bm u'| < k_{rad}, \\
\label{eq:fcd3_2} k_{cx} |\bm u| < \pi,
\end{gather}
respectively. Eq. \ref{eq:fcd1} is a criterion that varies with the displacement and the background pattern, while Eq. \ref{eq:fcd2_2} and Eq. \ref{eq:fcd3_2} are criteria which vary only with the background pattern.
For example, in the $k$-space at $P_0$ in Figure \ref{fig:FCD}, $k_{cx}$ = $k_{1x}$ = $k_{rad}$ for the checkered pattern and $k_{cx}$ = $k_{2x}$ = 2$k_{rad}$ for the lattice grid pattern. 
If the criterion in Eq.\ref{eq:fcd1} is not satisfied, aliasing artifacts will occur; if the criterion in Eq.\ref{eq:fcd2_2} is not satisfied, overlap will occur; and if the criterion in Eq.\ref{eq:fcd3_2} is not satisfied, phase wrapping will occur \cite{Wildeman2018}.

\subsection{Estimation of the apparent displacement from the pressure measured by the hydrophone}

\begin{figure}[t]
\begin{center}
\includegraphics[width=1.0\columnwidth]{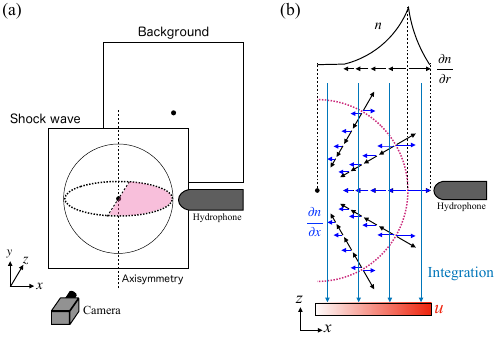}
\caption{Procedure of converting the pressure of underwater shock wave to the ``apparent displacement" $u$.  (a) Schematic of the BOS and hydrophone measurements of underwater shock waves in this study. The shock wave is axisymmetric about the direction of the laser beam (dotted straight line). (b) Constructed refractive index gradient field $\partial n/ \partial x$. To obtain $u$, integration of the $\partial n/ \partial x$ field was performed toward the $z$-direction in the pink area of (a).}
\label{fig:integ}
\end{center}
\end{figure}

The apparent displacement was estimated from the pressure measured by the hydrophone for comparison with FCD and PIV.
First, the density $\rho$ was obtained from the pressure using Tait equation: 
\begin{equation}
\rho = \left(\frac{p+B}{p_0+B}\right)^{\frac{1}{\alpha}} \rho_0,
\label{eq:tait}
\end{equation}
where the hydrostatic density $\rho_0$ is 998 kg/m$^3$, $B$ and $\alpha$ are 314 MPa and 7.15, respectively.
Then, the refractive index gradient $\partial n/ \partial r$ was calculated by differentiating the refractive index $n$, which was obtained from Eq. \ref{eq:gladstone} using the density $\rho$.
Because the shape and pressure of a laser-induced underwater shock wave are axisymmetric in the direction of the laser beam \cite{tagawa2016pressure,hayasaka2017effects}, the axisymmetric field of the refractive index gradient $\partial n/ \partial x$, as shown in Figure \ref{fig:integ}, was reconstructed. 
Finally based on Eq. \ref{eq:v_n}, the apparent displacement $u$ was obtained by integrating $\partial n/ \partial r$ along $z$-axis as shown in Fig. \ref{fig:integ}.

\section{Experiments}
\subsection{Experimental set-up}

\begin{figure}[t]
\begin{center}
\includegraphics[width=1.0\columnwidth]{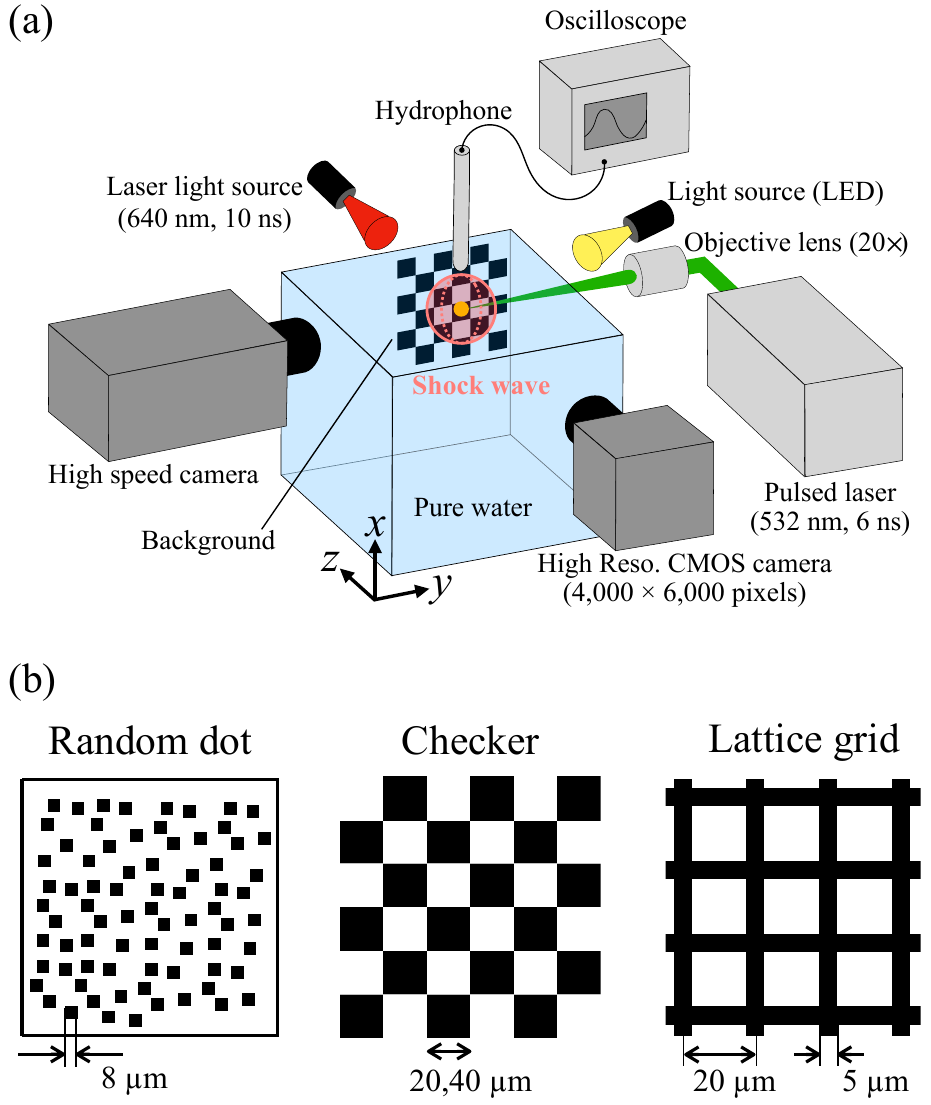}
\caption{(a) Experimental setup. (b) Background patterns. In the case of the random dot pattern, the dot size is 8 µm $\times$ 8 µm. 
In the case of the checkered pattern, the area of each tile is 20 \si{\um} $\times$ 20 \si{\um} or 40 \si{\um} $\times$ 40 \si{\um}. In the case of the lattice grid pattern, the width of the black line is 5 \si{\um}, and the area of the white region is 15 \si{\um} $\times$ 15 \si{\um}.}
\label{fig:setup}
\end{center}
\end{figure}
A schematic of the experimental setup is shown in Figure \ref{fig:setup}.
Three types of background patterns were used: random dots, checkered, and lattice grid.
To generate the underwater shock wave, a pulsed laser (Nd: YAG laser Nano S PIV, Litron Lasers Ltd., wavelength: 532 nm, pulse width: 6 ns) was focused through an objective lens (SLMPLN 20$\times$, Olympus, magnification: 20X, NA value: 0.25) into a point inside a container (100 mm $\times$ 100 mm $\times$ 100 mm) filled with ultrapure water.
The background was placed behind the shock wave (distance $Z_D$ about 0.7 - 2.5 mm).
A high-resolution CMOS camera (EOS 80D, Canon, resolution: 4,000 $\times$ 6,000 pixels, spatial resolution: 0.6 - 1.5 \si{\um}/pixel in this set-up) was used to capture the background images with and without the underwater shock wave.
The exposure time of the camera was set at approximately 0.5 s, and the timing of filming was adjusted by changing the timing of the light source.
A delay generator (Model 575, BNC) was utilized to synchronize the pulsed laser and laser light sources (SI-LUX 640, Specialized Imaging Ltd., wavelength: 640 nm, pulse width: 10-20 ns).
The shock wave pressure was measured using a hydrophone (Muller-Platte Needle Probe, Muller Instruments, Germany, effective diameter $<$ 0.25 mm).

To determine the apparent displacement on the background, an open-source PIV code (PIVlab, MATLAB, Thielicke and Stamhuis \cite{Thielicke2014}) and an open-source FCD code \cite{Wildeman2018} were used. 
A dot occupies about 5.5 - 11.3 pixel on the image. 
The PIV setting has a 50\% overlap with the fast Fourier transform (FFT) multipass interrogation (from 64 to 8 pixels). 

\begin{figure}[ht]
\begin{center}
\includegraphics[width=1.0\columnwidth]{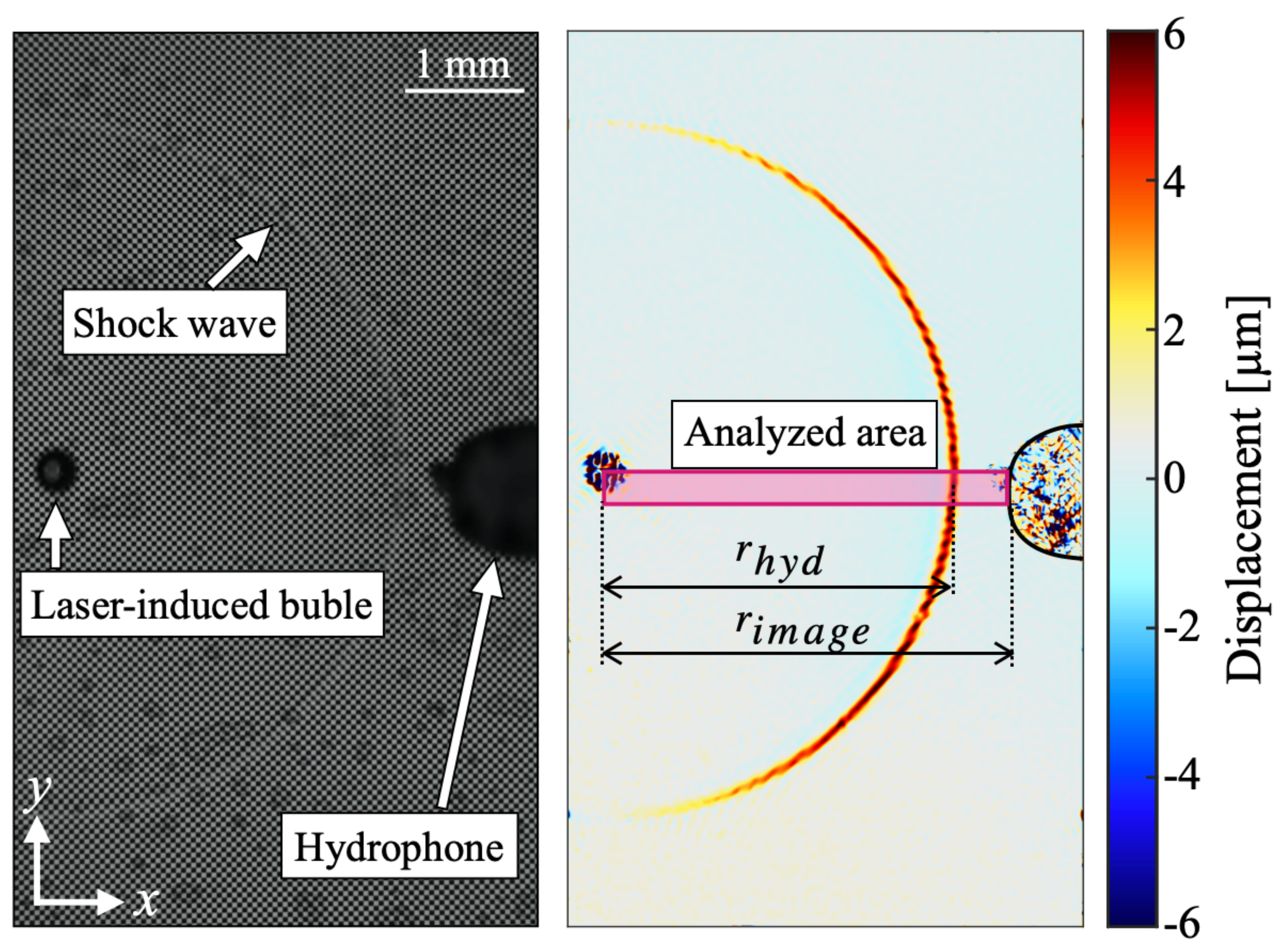}
\caption{A typical background image with a shock wave (left). $x$-components of the displacement field detected by FCD (right). The analyzed area was determined by considering the effective diameter of the hydrophone.}
\label{fig:shock}
\end{center}
\end{figure}

To compare with the displacements estimated by the hydrophone measurement, the displacements obtained by the BOS in the range of $\pm$0.125 mm in the $y$-direction from the center-line of the hydrophone were averaged along the $x$-axis (see Figure \ref{fig:shock}).
Note that the positions of the shock wave measured by the BOS and hydrophone in $x$-direction are slightly different, as shown in Figure \ref{fig:shock}. 
The pressure is inversely proportional to the distance of the shock wave propagation in this experimental range \cite{Vogel1996}.
Thus, for direct comparison with BOS results, the shock wave pressure measured by hydrophone was corrected using following equation:
\begin{equation}
p_{image} = \frac{r_{hyd}}{r_{image}} p_{hyd},
\label{eq:p_corr}
\end{equation}
where $p_{image}$ is the pressure at the time of imaging, $p_{hyd}$ is the pressure measured by the hydrophone, $r_{image}$ is the radius of the shock wave at the time of imaging, and $r_{hyd}$ is the distance between the center of the shock wave and the tip of the hydrophone.

\begin{figure}[ht]
\begin{center}
\includegraphics[width=1.0\columnwidth]{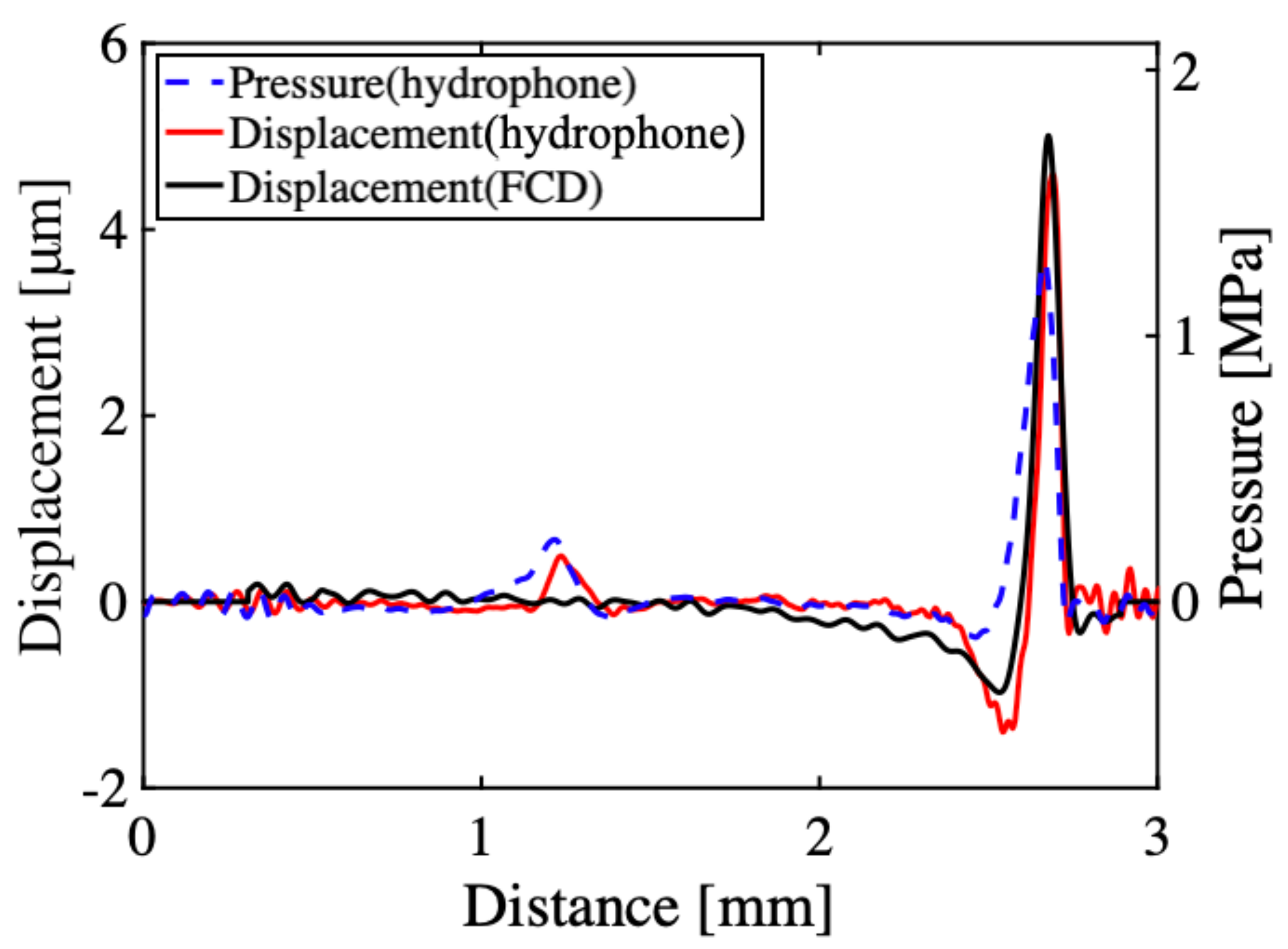}
\caption{The apparent displacement in the $x$-direction and shock wave pressure as a function of the distance from the center of the shock wave.}
\label{fig:disp}
\end{center}
\end{figure}

In Figure \ref{fig:disp}, the results of the FCD and the displacement calculated from the pressure measured by hydrophone are shown.
In following sections, we focus on the largest apparent displacement at the distance near 2.7 mm (first peak).
Note that a second peak (distance around 1.2 mm) appears in the estimated displacement because a part of the shock wave was reflected on the background.
The second peak in the image analysis does not appear because the integral of $\partial n / \partial  x$ is very small.


\subsection{Experimental results}

\begin{figure}[t]
\begin{center}
\includegraphics[width=1.0\columnwidth]{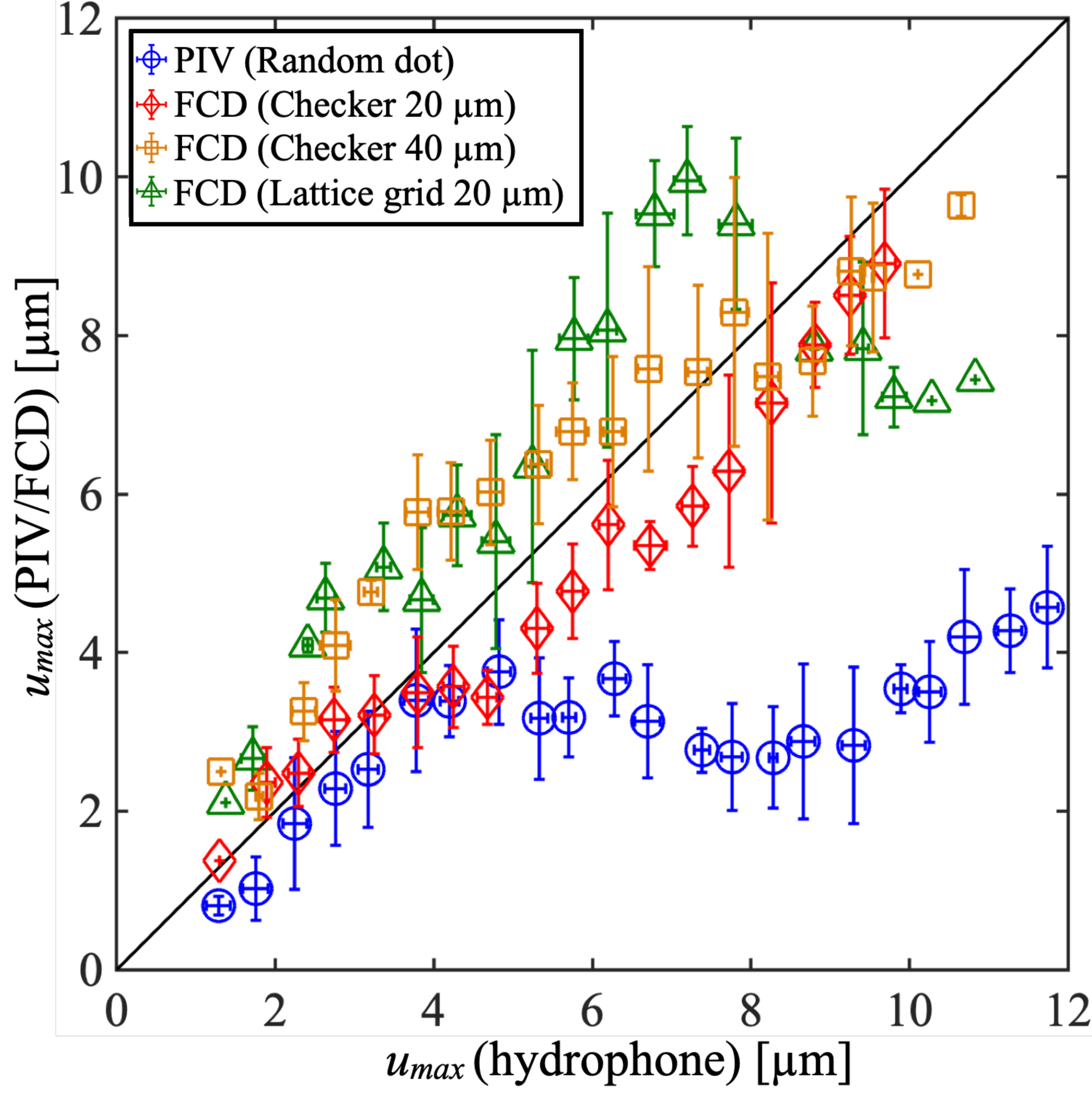}
\caption{The maximum displacement estimated from the pressure measured with hydrophone $u_{max}$(hydrophone) and the maximum displacement detected by PIV/FCD $u_{max}$(PIV/FCD). Error bars indicate the standard deviations.}
\label{fig:umax}
\end{center}
\end{figure}

\begin{figure}
\begin{center}
\includegraphics[width=1.0\columnwidth]{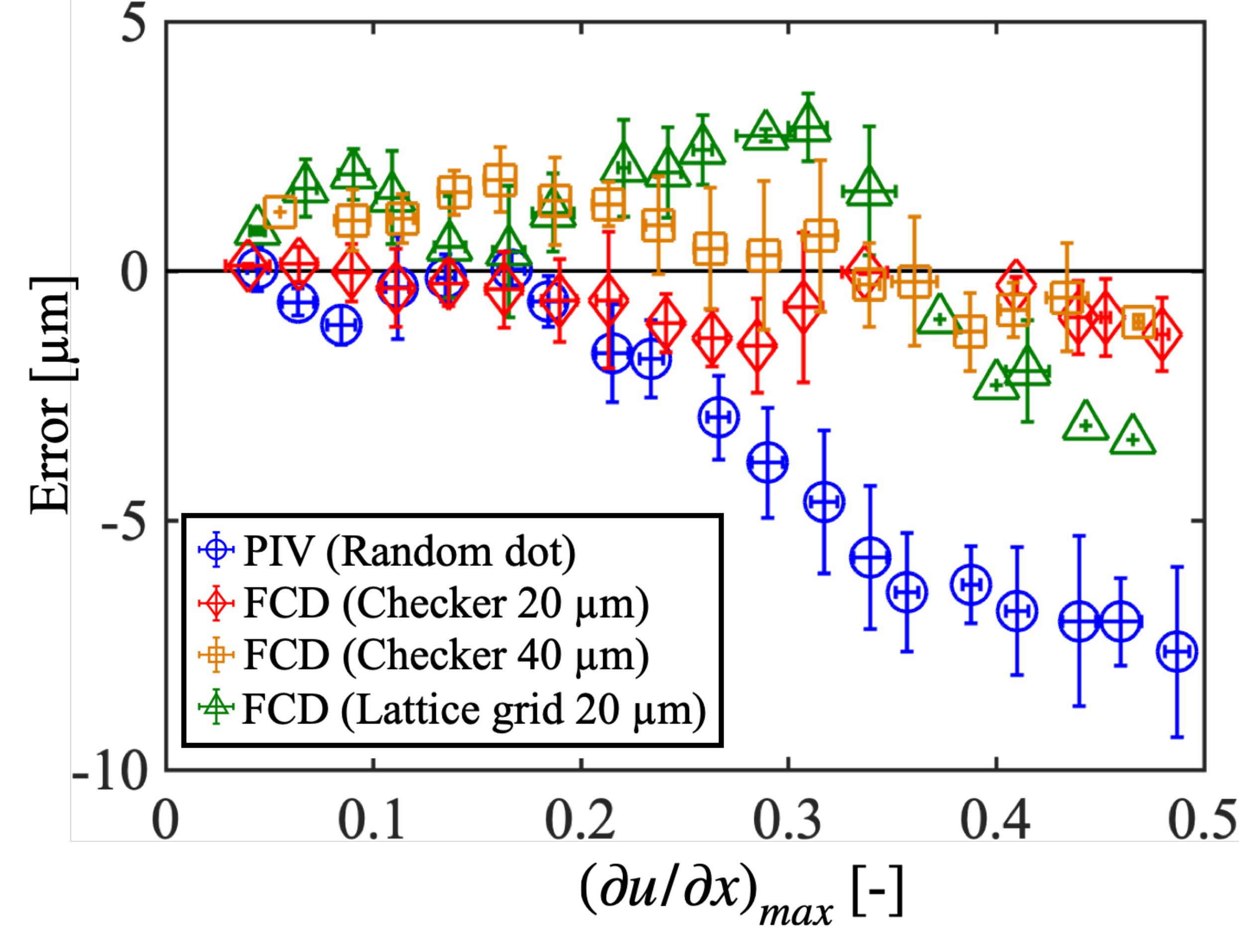}
\caption{The error which is the difference between the maximum displacement estimated from the pressure $u_{max}$(hydrophone) and the maximum displacement detected by PIV/FCD $u_{max}$(PIV/FCD) against the maximum displacement gradient of the displacement estimated from the pressure measured by hydrophone $(\partial u/ \partial x)_{max}$. Error bars indicate the standard deviations.}
\label{fig:ug}
\end{center}
\end{figure}

The maximum displacement detected by PIV or FCD, $u_{max}$(PIV/FCD) and maximum displacements calculated from the pressure, $u_{max}$(hydrophone), are shown in Figure \ref{fig:umax}.
For $u_{max}$(hydrophone) $>$ 4 \si{\um}, the maximum displacement detected by PIV is about 4 \si{\um}, which is significantly lower than that estimated from the measurements of the hydrophone.
Such underestimation is due to a significant distortion of the shape of the dots when $u_{max}$(hydrophone) $>$ 4 \si{\um}, making it difficult for PIV to accurately detect the displacement.
In contrast, the displacements obtained by the FCD reasonably agrees with the estimated displacement.

The error of the PIV and FCD methods is defined as the difference between the maximum displacement detected using each method and that estimated from the pressure measured using hydrophone.
In Figure \ref{fig:ug}, the errors are plotted against the maximum displacement gradient $(\partial u/ \partial x)_{max}$, which represents the magnitude of the distortion.
As shown in the figure, PIV underestimates the displacement when the maximum displacement gradient is greater than 0.20.
This is because the displacement detected by the cross-correlation method does not match the original displacement owing to the large distortion of the dots as discussed in Section 1.
On the other hand, FCD can reasonably detect the maximum displacement for the displacement gradient $\partial u/ \partial x < 0.5$, which is approximately 2.5 times larger than 0.20, the measurement range of PIV-BOS.
Among the FCD methods, the error of 20 \si{\um} checkers is less than the error of 40 \si{\um} checkers and 20 \si{\um} lattice grid. 
Besides, it is also observed that 20 \si{\um} lattice grid, the error shifts from overestimation to underestimation at the displacement gradient between 0.3 and 0.4.
To analyze the reasons for such trends, we utilized synthetic images to investigate the applicable range for displacement detection of FCD with various background patterns.
Synthetic images were used because the images captured from the experiment might be affected by random errors such as image blurring and slight misalignment of the experimental setup.
The analysis are discussed in Chapter 4.



\section{Numerical approach: measurement limit of FCD-BOS}
\subsection{Numerical condition}

\begin{figure*}[!ht]
\begin{center}
\includegraphics[width=1.8\columnwidth]{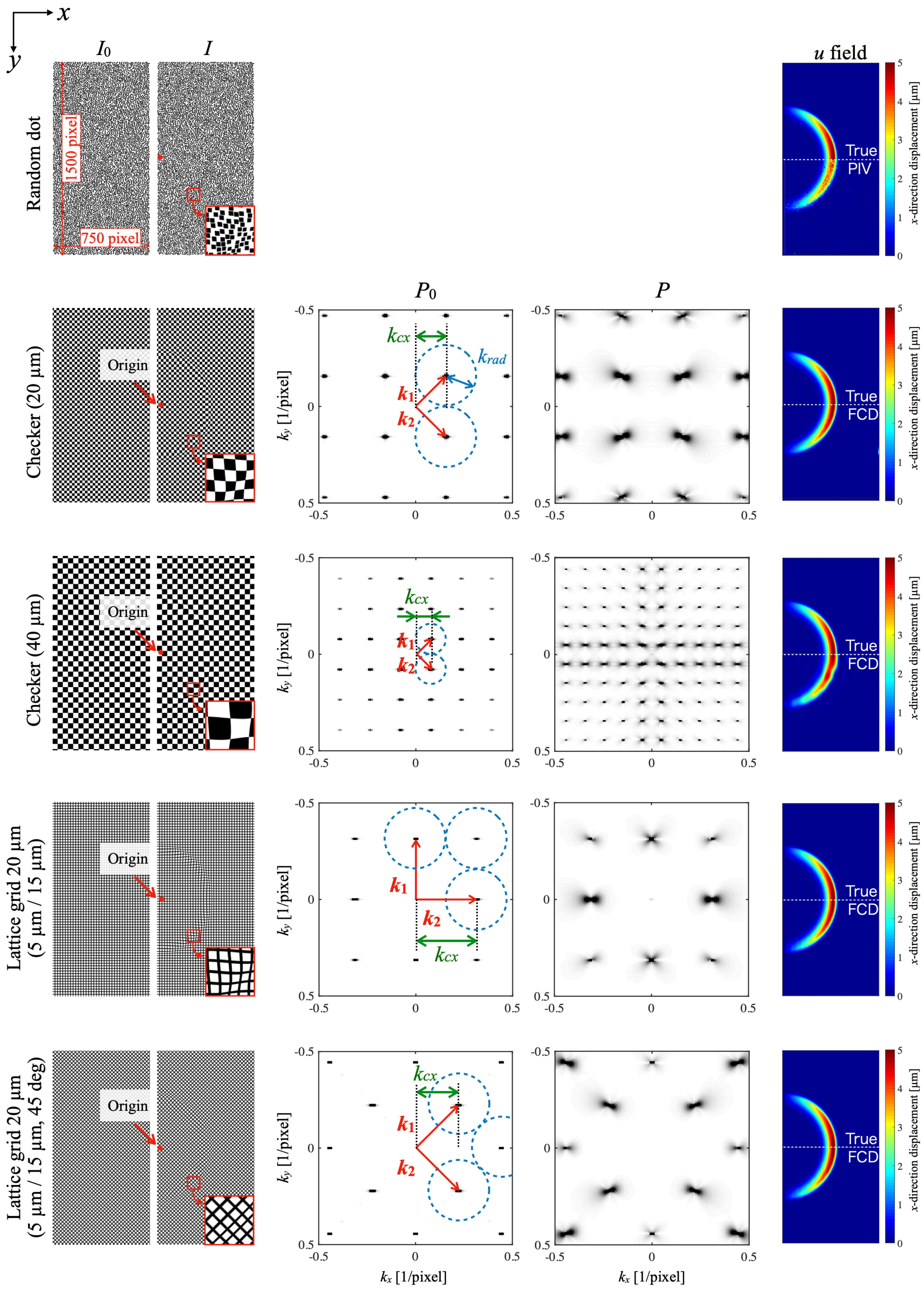}
\caption{Synthetic images (1500 $\times$ 750 pixels) and the $x$-direction displacement $u$ (maximum displacement is 5 \si{\um} (= 5 pixels) in this figure). The radius of the circle in each $P_0$ is $k_{rad}$. Symbols are explained in the caption of Figure \ref{fig:FCD}.}
\label{fig:numeri}
\end{center}
\end{figure*}

In order to determine the applicable range of displacement detection of FCD, we used synthetic images (image size: 1500 $\times$ 750 pixels, spatial resolution: 1 \si{\um}/pixel) that imitate three types of background patterns: random dots, lattice grid, and checkered.
The synthetic images were modulated with synthesized apparent displacement $u(r)$, i.e., ``true displacement", based on the Gaussian distribution as follows:
\begin{equation}
u(r) = 
\left\{
\begin{array}{ll}
u_{max} \exp \left(- \dfrac{(r-r_c)^2}{\beta} \right) & \quad (r \geq r_c)  \vspace{1em} \\
u_{max} \exp \left(- \dfrac{(r-r_c)^2}{9 \beta} \right) & \quad (r < r_c),
\end{array}
\right.
\label{eq:numeri_u}
\end{equation}
where $u_{max}$ varies between 1 - 11 \si{\um} at an interval of 1 \si{\um}, $r$ is the distance from the origin indicated in Figure \ref{fig:numeri}, ${r_c}$ is set to 400 \si{\um}, and $\beta$ is set to $400 \times 10^{-12}$ m$^2$.
This $u(r)$ resembles the displacement of a shock wave. 
For each interval of $u_{max}$, the origin of $u(r)$ was issued to 20 different positions relative to the background pattern. 
These 20 data were used to obtain the averaged value and its error range.
\subsection{Numerical results}
\subsubsection{Size and orientation of the background patterns}
\begin{figure}[t]
\begin{center}
\includegraphics[width=0.9\columnwidth]{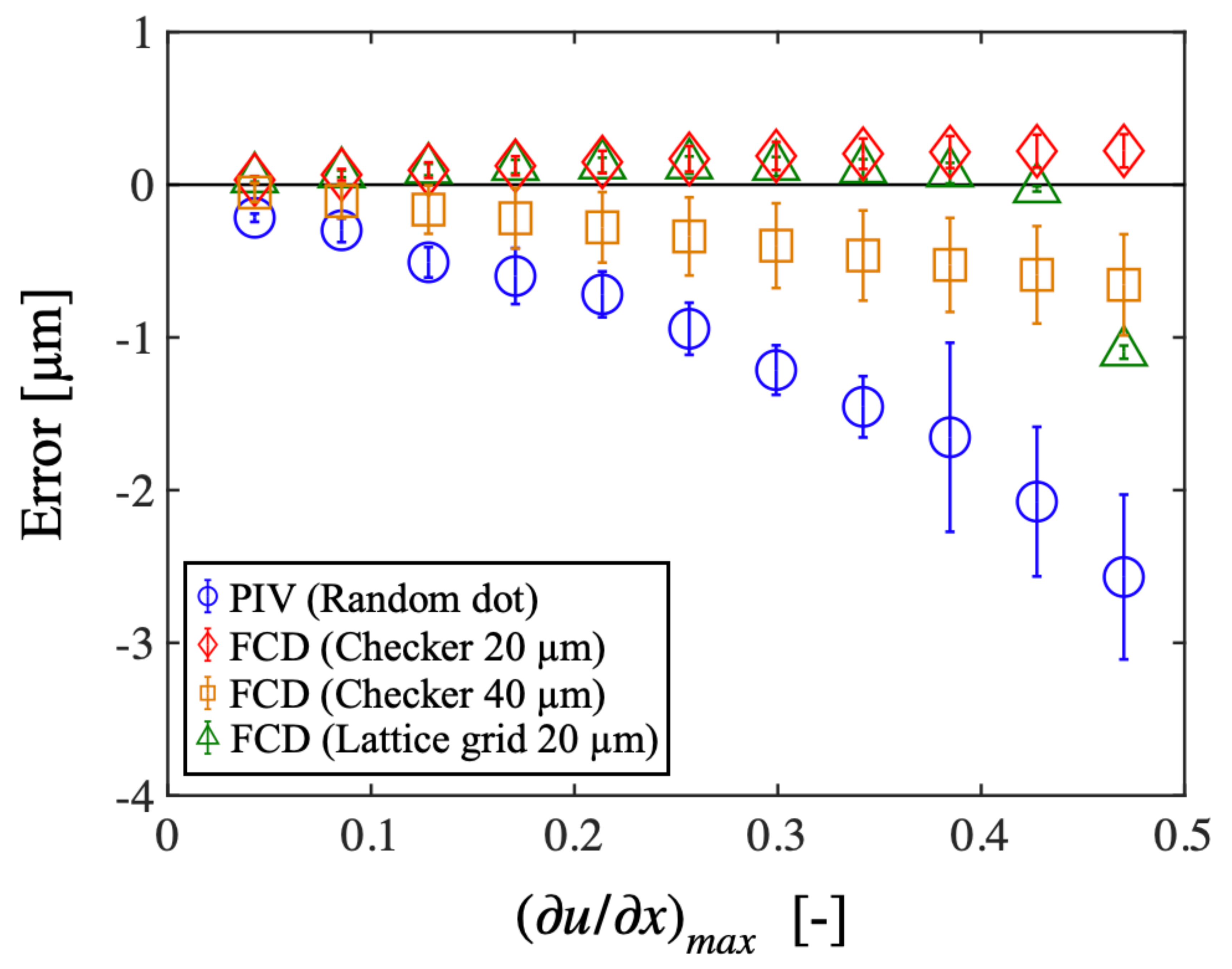}
\caption{The maximum displacement gradient of the displacement given modulated to the reference image $(\partial u/ \partial x)_{max}$ and the error which is the difference between the maximum displacement given modulated to the reference image $u_{max}$(estimation) and the maximum displacement detected by PIV/FCD $u_{max}$(PIV/FCD). Error bars indicate the standard deviations estimated by using 20 sets of images.}
\label{fig:ug_numeri}
\end{center}
\end{figure}

\begin{figure}[!t]
\begin{center}
\includegraphics[width=0.95\columnwidth]{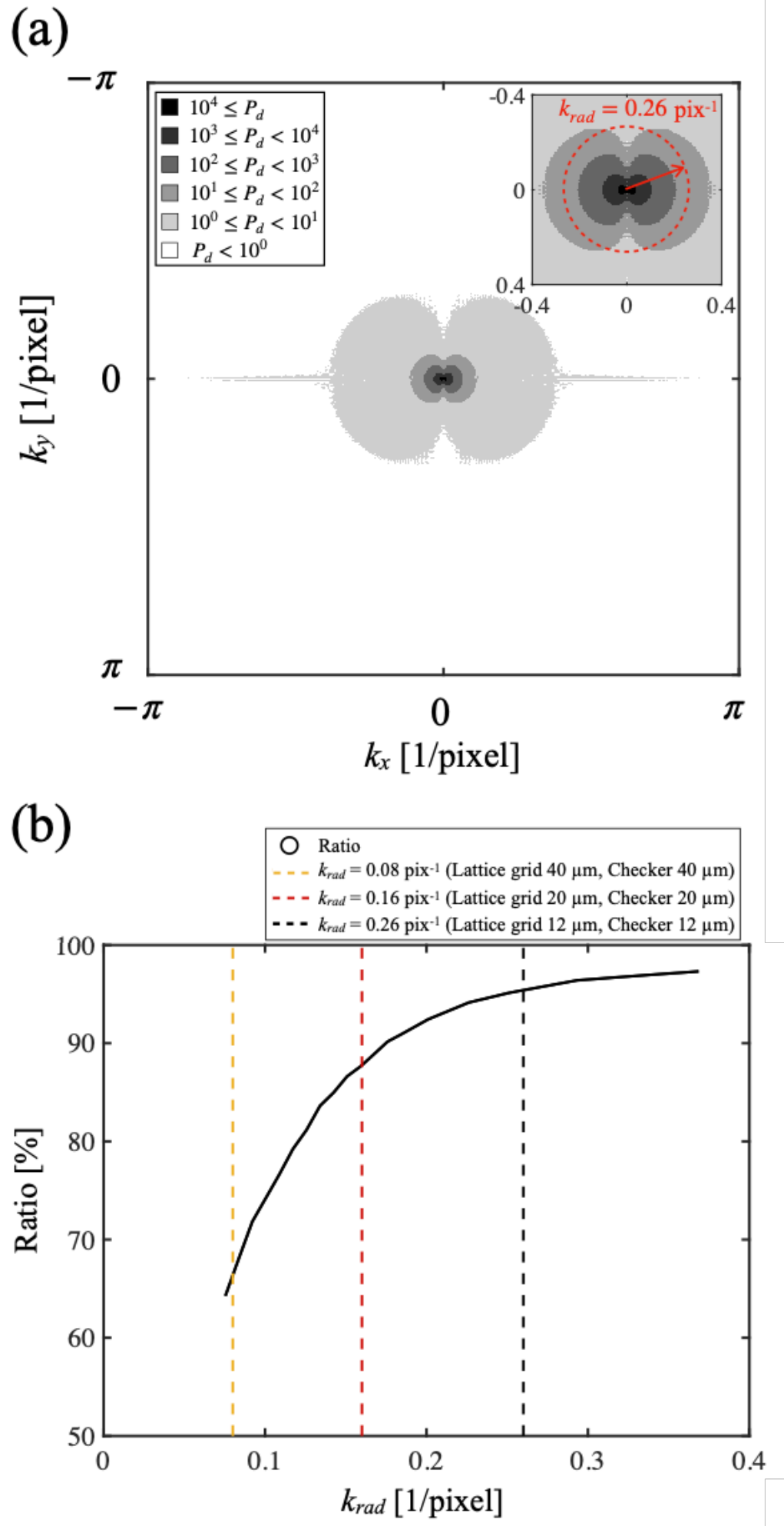}
\caption{(a) Power spectra $P_d$, which is obtained by the Fourier transformation of the true displacement $u(r)$, for $u_{max} =$ 1 \si{\um}.
The subfigure on the upper right shows the enlarged view. The radius of the circle, $k_{rad}$, varies depending on the pattern. (b) The ratio of the sum of $P_d$ contained within $k_{rad}$ to the total sum of $P_d$.}
\label{fig:ks_ratio}
\end{center}
\end{figure}

\begin{figure}[ht]
\begin{center}
\includegraphics[width=1.0\columnwidth]{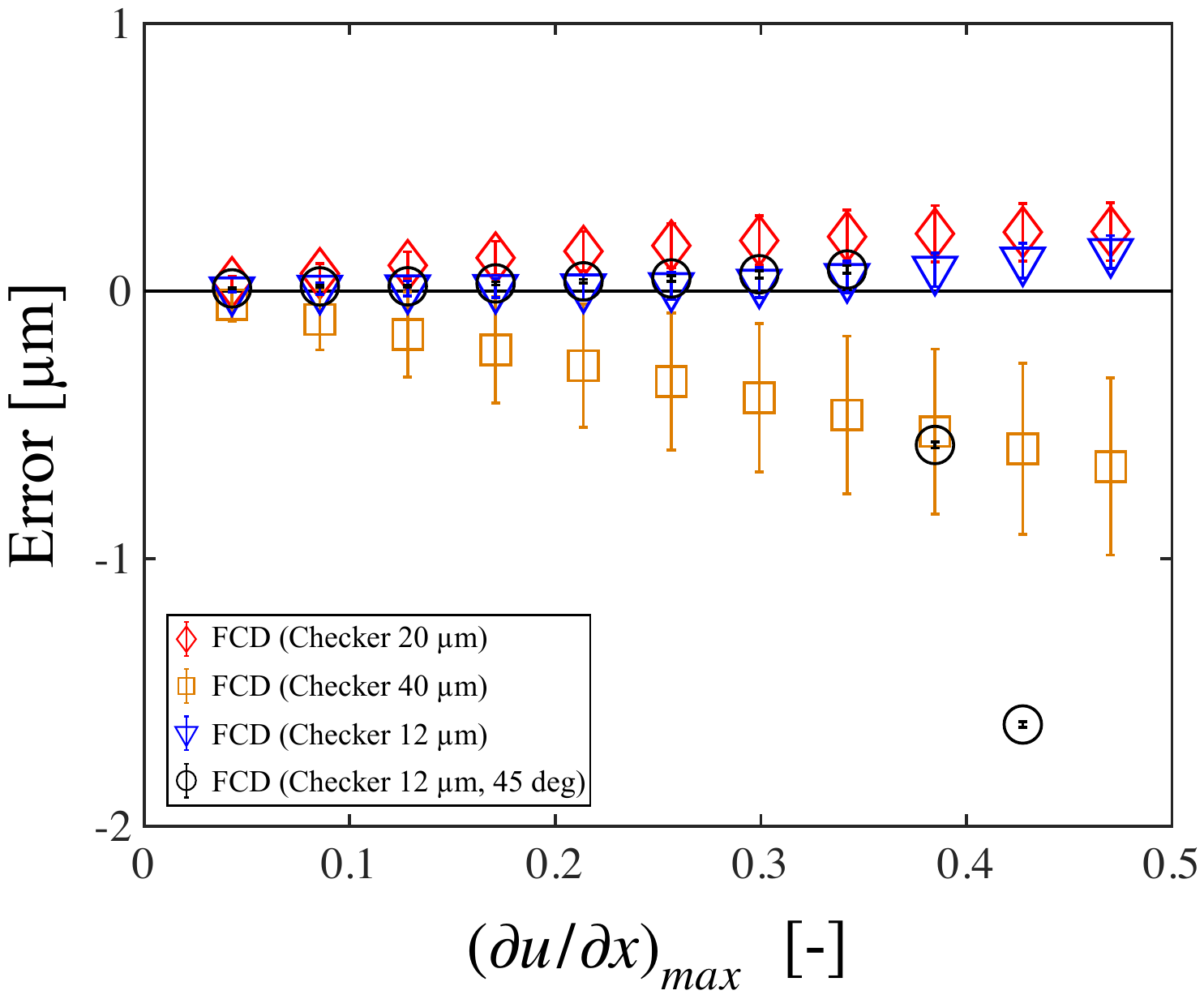}
\caption{The measurement error of checkered patterns of different sizes and orientations against the maximum displacement gradient. Error bars indicate the standard deviations.}
\label{fig:ug_checker}
\end{center}
\end{figure}

\begin{figure}[ht]
\begin{center}
\includegraphics[width=1.0\columnwidth]{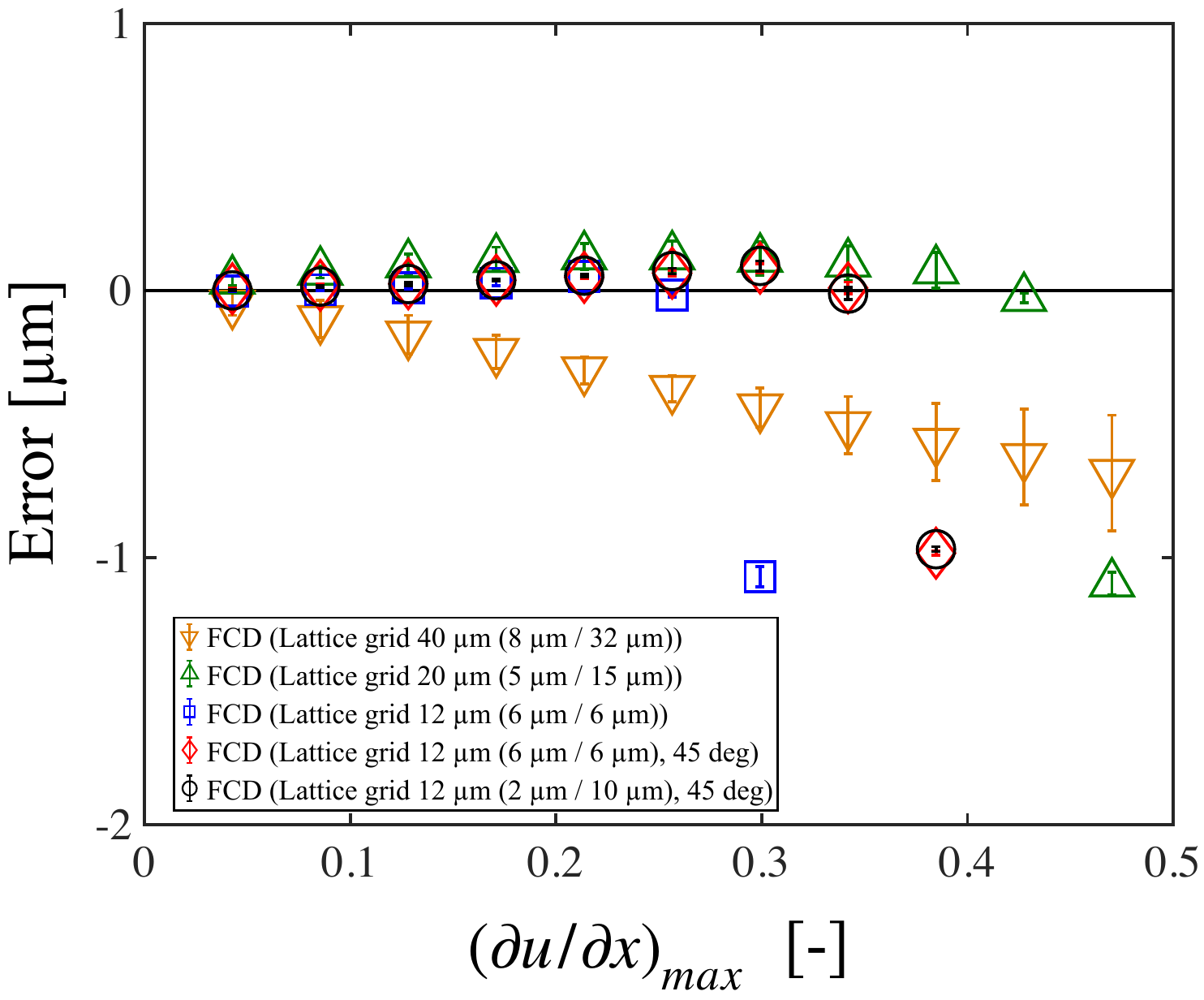}
\caption{The measurement error of lattice grid patterns of different sizes and orientations against the maximum displacement gradient. Error bars indicate the standard deviations.}
\label{fig:ug_grid}
\end{center}
\end{figure}
\begin{figure*}[!ht]
\begin{center}
\includegraphics[width=2.0\columnwidth]{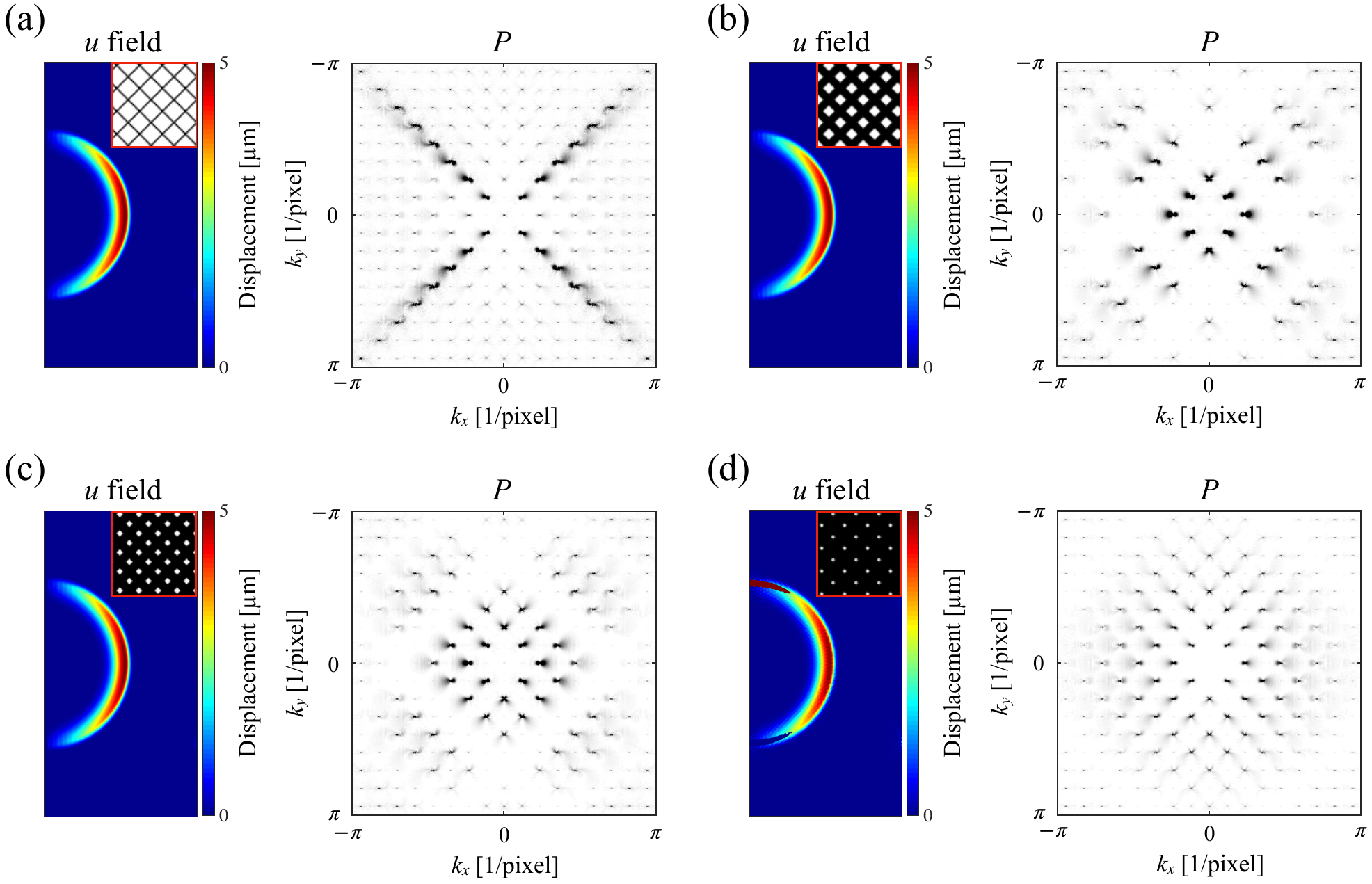}
\caption{The displacement fields and the wavenumber intensity of synthetic images modulated with true displacement for $u_{max} =$ 5 \si{\um} measured using 12 \si{\um} lattice grid patterns of different line thickness:
(a) 1 \si{\um}/11 \si{\um}, (b) 6 \si{\um}/6 \si{\um},(c) 8 \si{\um}/4 \si{\um}, and (d) 10 \si{\um}/2 \si{\um}.
The lattice grid patterns are shown in the insets.
}
\label{fig:grid_thick}
\end{center}
\end{figure*}

Figure \ref{fig:numeri} shows the typical result for $u_{max} =$ 5 \si{\um}.
Overall, the results show similar trends as the experimental results.
The displacement detected by PIV is noisy, while those detected by the FCD methods are similar to the true displacement.
However, the measurement accuracy of FCD depends on the background pattern where the displacement fields of 20 \si{\um} checkers, 20 \si{\um} ($0^\circ$) lattice grid, and 20 \si{\um} ($45^\circ$) lattice grid are almost identical to the true data, but the displacement field of 40 \si{\um} checkers is partially different from true data.
The reason for this is discussed in the later part of this section.
For the lattice grid patterns without rotation ($0^\circ$) and with a rotation of $45^\circ$, $k_{cx}$ are 0.32 pixel$^{-1}$ and 0.23 pixel$^{-1}$, respectively although $k_{rad}$ are 0.16 pixel$^{-1}$ for both patterns.
Nevertheless, since the lattice grid patterns of both $0^\circ$ and $45^\circ$ satisfy Eq. \ref{eq:fcd3_2}, there is no notable difference between their displacement fields.

Relations between the maximum displacement gradient and error are shown in Figure \ref{fig:ug_numeri}.
PIV underestimates the apparent displacement as the displacement gradient increases, similar to the experimental results shown in Figure \ref{fig:ug}.
On the other hand, the measurement error of FCD (20 \si{\um} checkers) is sufficiently low for $(\partial u/ \partial x )_{max} < 0.5$ while that of FCD (20 \si{\um} lattice grid) has non-trivial values for $(\partial u/ \partial x )_{max}>0.45$.
Since the value of $k_{cx}$ for the lattice grid pattern (20 \si{\um}) is 0.32 pixel$^{-1}$, based on Eq. \ref{eq:fcd3_2}, phase wrapping occurs when the displacement is larger than 9.8 \si{\um}, which corresponds to a displacement gradient of 0.42.
Thus, FCD (20 \si{\um} lattice grid) underestimates the displacement when $(\partial u/ \partial x )_{max}$ $>$ 0.45.
The FCD (40 \si{\um} checkers) gradually underestimates the displacement as the displacement gradient increases because of the resulting aliasing artifacts as described by Eq. \ref{eq:fcd1}.

To investigate further how the criterion in Eq.\ref{eq:fcd1} offsets the results in our condition, the power spectra $P_d$, which is obtained by the Fourier transformation of the true displacement $u(r)$, is analyzed.
Since the analysis results are similar for every displacement between $u_{max}$= 1 and 11 \si{\um}, here only the results for the synthetic image of $u_{max}$= 1 \si{\um} are discussed.
Figure \ref{fig:ks_ratio}(a) shows the $P_d$ for $u_{max}$= 1 \si{\um}, where the red dashed circle in the zoomed-in figure shows the boundary of $k_{rad} =$ 0.26~pixel$^{-1}$. 
Since Eq.\ref{eq:fcd1} shows that the measurement accuracy correlates to a ratio of the sum of $P_d$ contained within $k_{rad}$ to the total sum of $P_d$, the ratio is plotted as a function of $k_{rad}$ in Figure \ref{fig:ks_ratio}(b).
In the same figure, dashed vertical lines are plotted for the $k_{rad}$ values of the patterns of different sizes, i.e., $k_{rad} =$ 0.08~pixel$^{-1}$ (40 \si{\um} checkers and lattice grid), 0.16~pixel$^{-1}$ (20 \si{\um} checkers and lattice grid), and 0.26~pixel$^{-1}$ (12 \si{\um} checkers and lattice grid).
The ratios for $k_{rad} =$ 0.08~pixel$^{-1}$, 0.16~pixel$^{-1}$, and 0.26~pixel$^{-1}$ are 65\%, 87\%, and 95\%, respectively.
Since ratio for 40 \si{\um} checkers is lower that that for 20 \si{\um} checkers, the resulting aliasing artifacts in 40 \si{\um} checkers have caused more noise in the detected displacements than those of 20 \si{\um} checkers, as shown in Figure \ref{fig:numeri}.
Besides, they have also caused the underestimation of the displacement by 40 \si{\um} as shown in Figure \ref{fig:ug_numeri}.
Such results show that the measurement accuracy can be further improved when the ratio is increased through increasing the values of $k_{rad}$ using the patterns of smaller size.

Here, we focus on $k_{rad} =$ 0.26~pixel$^{-1}$ (12 \si{\um} checkers and lattice grid) which correspond to a ratio of 95\%.
For that, the measurement accuracy of 12 \si{\um} checkers and lattice grid were evaluated based on the respective measurement errors shown in Figure \ref{fig:ug_checker} and Figure \ref{fig:ug_grid}.

In Figure \ref{fig:ug_checker}, when the maximum displacement gradient is less than 0.35, 12 \si{\um} checkers has the smallest error. 
Similarly, in Figure \ref{fig:ug_grid}, when the maximum value of the gradient is less than 0.25, 12 \si{\um} lattice grid has the smallest error.
However, in Figure \ref{fig:ug_checker}, the errors of 12 \si{\um} and 12 \si{\um} 45$^\circ$ checkers deviate from one another when $(\partial u/\partial x)_{max}>$ 0.35.
Similarly, in Figure \ref{fig:ug_grid}, the errors of 12 \si{\um} and 12 \si{\um} 45$^\circ$ lattice grids also deviate from one another when $(\partial u/\partial x)_{max}>$ 0.25.
The reason for these deviations can be understood based on Eq. \ref{eq:fcd3_2}.
For checkered pattern of 12 \si{\um} $45^\circ$ ($k_{cx}$ = 0.37 pixel$^{-1}$), the maximum measurable displacement is 8.5 \si{\um}, which corresponds to $(\partial u/\partial x)_{max}$= 0.37, while for that of 12 \si{\um} $0^\circ$ ($k_{cx}$ = 0.26 pixel$^{-1}$), the maximum measurable displacement is 12 \si{\um}, which corresponds to $(\partial u/\partial x)_{max}$= 0.51.
Thus, as shown in Figure \ref{fig:ug_checker}, the measurement errors of 12 \si{\um} 45$^\circ$ checkers show non-trivial values for $(\partial u/\partial x)_{max}>$ 0.35, owing to phase wrapping.

As for the results for lattice grid patterns shown in Figure \ref{fig:ug_grid}, phase wrapping occurs for 12 \si{\um} $0^\circ$ and 12 \si{\um} $45^\circ$ when $(\partial u/\partial x)_{max} > 0.26$ and $0.34$, respectively.
Based on Eq. \ref{eq:fcd3_2}, for the lattice grid pattern of 12 \si{\um} $45^\circ$ ($k_{cx}$ = 0.37 pixel$^{-1}$), the maximum measurable displacement is 8.5 \si{\um}, which corresponds to $(\partial u/\partial x)_{max}$= 0.37, while for that of 12 \si{\um} $0^\circ$ ($k_{cx}$ = 0.52 pixel$^{-1}$), the maximum measurable displacement is 6.0 \si{\um}, which corresponds to $(\partial u/\partial x)_{max}$= 0.26.
Therefore, phase wrapping started occurring at a smaller value of $(\partial u/\partial x)_{max} $ for 12 \si{\um} of $0^\circ$.
As for the lattice grids of 12 \si{\um} (2\si{\um}/10\si{\um}) $45^\circ$ and 12 \si{\um} (6\si{\um}/6\si{\um}) $45^\circ$, phase wrapping occurred at the same value of $(\partial u/\partial x)_{max}$ because $k_{cx}$ is identical.

In summary, the limiting factors for FCD can be understood from the parameters in Fourier space as shown in Eq. \ref{eq:fcd1}, Eq. \ref{eq:fcd2_2}, and Eq. \ref{eq:fcd3_2}. 
It is worth noting from a practical view point that the results showed that FCD of lattice grid patterns can measure displacement as accurate as that of checkered patterns, although it is slightly more restricted.

\subsubsection{Line thickness of lattice grid pattern}

For practical purpose, we investigated the influence of line thickness of the lattice grid pattern.
The results of the different line thicknesses of 12 \si{\um} lattice grid patterns for the maximum displacement $u_{max}$ = 5 \si{\um} are shown in Figure \ref{fig:grid_thick}.
In these cases, all FCD criteria (Eq. \ref{eq:fcd1}, Eq. \ref{eq:fcd2_2}, and Eq. \ref{eq:fcd3_2}) are satisfied.
While $k_{cx}$ and $k_{rad}$ are the same for all cases, the wavenumber intensity $P$ of the synthetic images depends on the line thickness.
If the lines are significantly thicker than the white regions, as shown in Figure \ref{fig:grid_thick}(d), the $u$ field is different from the true displacement described in Eq. \ref{eq:numeri_u}. 
Since $u$ fields in Figure \ref{fig:grid_thick}(a)(b) are identical to true displacements, it is recommended to use lattice grid patterns with lines that are thinner than the white regions for FCD-BOS measurements.


\section{Conclusion}
We applied Fast Checkerboard Demodulation (FCD) method for Background Oriented Schlieren (BOS) technique.
In general, BOS tecnique uses a random-dot background to detect the displacement.
However, to measure large density-gradient fluid fields such as shock waves, the dotted pattern distorts causing the displacement detection by cross-correlation PIV method fails to work properly.
From the experimental results, FCD is superior to PIV in terms of detecting large displacement gradients and the maximum measurable displacement gradient of FCD is 2.5 times larger than that of PIV.
Through analysis using synthetic data, the limiting factors of the FCD can be understood from the parameters in Fourier space as shown in Eq.\ref{eq:fcd1}, Eq.\ref{eq:fcd2_2}, and Eq.\ref{eq:fcd3_2}.
The results of the analysis shows that line thickness influences the limiting factors of the FCD of lattice grid patterns.
On a final note, FCD-BOS can be easily constructed and customized for various thermal fluids since it just consists of a commercially available grid scale, still camera and light source.
Based on measurement limitation of FCD-BOS reported in this paper, users of FCD-BOS could select proper background with suitable parameters such as the line width of lattice grid patterns and region size of checker patterns.






\section*{Acknowledgement}
This work was supported by the Kawai Foundation for Sound Technology $\&$ Music, KAKENHI Grant-in-Aid for Scientific Research (A), Grant Number 20H00222/20H00223, and JST, PRESTO Grant Number JPMJPR21O5, Japan.
The authors would like to thank Masaharu Kameda (Professor, Tokyo University of Agriculture and Technology) and Jingzu Yee (PhD candidate, Tokyo University of Agriculture and Technology) for their valuable discussions and suggestions.

\bibliographystyle{ieeetr}
\bibliography{ref}
\end{document}